\documentstyle[12pt]{article}

\begin{document}

\title{\Large {\bf Euler numbers of four-dimensional rotating black holes
            with the Euclidean signature } }

\author{{{\large Zheng Ze Ma}} \thanks{Electronic address: 
           z.z.ma@seu.edu.cn} 
  \\  \\ {\normalsize {\sl Department of Physics, Southeast University, 
         Nanjing, 210096, P. R. China } }}

\date{}

\maketitle

\begin{abstract}

\indent

  For a black hole's spacetime manifold in the Euclidean signature, 
its metric is positive definite and therefore a Riemannian manifold.
It can be regarded as a gravitational instanton and a topological 
characteristic which is the Euler number is associated. 
In this paper we derive a formula for the Euler numbers of 
four-dimensional rotating black holes by the integral of the Euler 
density on the spacetime manifolds of black holes. Using this 
formula, we obtain that the Euler numbers of Kerr and Kerr-Newman 
black holes are $2$. We also obtain that the Euler number of the 
Kerr-Sen metric in the heterotic string theory with one boost 
angle nonzero is $2$ that is in accordence with its topology. 

\end{abstract}

~~~  PACS number(s): 04.20.Gz, 04.50.+h, 04.70.-s, 11.25.Mj

\section{Introduction}

\indent

  Euler number is one of the topological invariants for Riemannian 
manifolds. It is the alternative sum of the Betti numbers, i.e.,
$\chi=\sum(-1)^{p}B_{p}$.  For a black hole's spacetime manifold in 
the Euclidean signature its metric is positive definite and therefore
is a Riemannian manifold. Therefore the Euler characteristic is 
associated with the manifold. 
The topologies of Kerr and Kerr-Newman black holes in the metrics of the
maximal analytic extensions are $R^{2}\times S^{2}$. The Euler number of 
a product manifold $M=M_{1}\times M_{2}$ is the product of the Euler 
numbers of each manifold $\chi(M)=\chi(M_{1})\times \chi(M_{2})$. Thus 
one obtains that the Euler numbers for Kerr and Kerr-Newman black holes 
are $2$ as well as other four-dimensional rotating black holes, such as
those appear in the heterotic string theory [1-4]. 
The direct calculations for the Euler numbers of four-dimensional 
spherically symmetric black holes were given by the authors of 
Refs. [5-8]. The direct calculation for the Euler number of the Kerr 
metric was performed in Ref. [6]. However the direct calculation 
for the Euler number of the Kerr-Newman metric seems missing from the 
literature. In this paper we will derive a formula of Eq. (2.28) for 
the Euler numbers of four-dimensional rotating black holes using the 
Gauss-Bonnet formula. According to the formula (2.28), we can verify 
that the Euler numbers for the Kerr and Kerr-Newman metrics are $2$. 
We can also obtain that the Euler number for the Kerr-Sen metric [1] in 
the heterotic string theory given by Sen is $2$. Therefore the formula 
obtained in this paper for the Euler numbers of the four-dimensional 
black holes has some universality.

  In this section we first write down the Gauss-Bonnet formula 
that is used to calculate the Euler numbers. Let $M^{n}$ be a compact 
orientable Riemannian manifold of an even dimension $n$. According to 
Refs. [9,10] we define in $M^{n}$ the intrinsic exterior differential 
form $\Omega$ of degree $n$ which is equal to a scalar invariant of 
$M^{n}$ multiplied by the volume element 
$$
  \Omega=(-1)^{p-1}\frac{1}{2^{2p}\pi^{p}p!}
         \epsilon_{a_{1}\cdots a_{2p}}\Omega^{a_{1}a_{2}}\wedge
         \cdots\wedge\Omega^{a_{2p-1}a_{2p}}~,
  \eqno{(1.1)}  $$ 
where $n=2p$, $\epsilon_{a_{1}\cdots a_{2p}}$ is a symbol which is equal
to $+1$ or $-1$ according to $i_{1}$, $\cdots$, $i_{n}$ to form an even 
or odd permutation of $1$,$\cdots$,$n$, and is zero otherwise. 
The unit vectors of the original manifold $M^{n}$ form a larger 
and higher dimensional manifold 
$M^{2n-1}$ of dimension $2n-1$. Chern [10] has shown that $\Omega$ is 
equal to the exterior derivative of a differential form $\Pi$ of degree 
$n-1$ in $M^{2n-1}$: 
$$
  \Omega=d\Pi~.
  \eqno{(1.2)}  $$ 
Chern has obtained that the Euler number of the manifold can be expressed 
by the integral of the $n$-form $\Omega$ on the manifold:
$$
  \chi=\int_{M}\Omega~.
  \eqno{(1.3)}  $$ 
Supposing that the definition of $\Pi$ in the manifold $M^{2n-1}$ can be 
extended to be defined in the original manifold $M^{n}$, then by using 
the Stokes theorem the integral of Eq. (1.3) can be converted to the 
integral of $\Pi$ on the boundaries of the manifold. (In fact such 
extensions of the definition of $\Pi$ in the original manifolds $M^{n}$ 
are exist in the practical calculations such as the calculation of 
the four-dimensional case proceeded in the following of this paper.) 
According to Chern the manifold may be the compact orientable Riemannian. 
For compact manifolds they include compact manifolds with boundaries and 
compact manifolds without boundaries usually. 
For a four-dimensional spherically symmetric or rotating black hole, 
the horizon or outer horizon is a null surface which divides the 
spacetime into two parts. For an ideally permanent black hole, any 
information inside the horizon cannot escape out of the horizon. 
Therefore for the observer outside the horizon the area inside the 
horizon is unphysical. The physical area can be regarded to be 
surrounded by two three-dimensional hypersurfaces, one is the horizon or 
outer horizon, the other lies at infinity. The later can be realized 
through mapping the points of constant $t$, $\theta$ and $\phi$ with 
$r\rightarrow\infty$ onto a three-dimensional hypersurface at infinity. 
These two boundaries make the physical area of a black hole's spacetime 
be a compact orientable manifold with two boundaries. In their metrics of 
Euclidean signatures they are Riemannian. Therefore we can apply the 
Gauss-Bonnet formula (1.3) directly to calculate their Euler numbers 
and the integral areas can only be taken outside the horizons. 
\footnote{The reason why the integral area can only be taken outside 
  the horizon of a black hole can also be understood from the point of
  combinatorial topology. From the sense of combinatorial topology, 
  one divides a compact orientable manifold into many cells. The Euler 
  number of the manifold is defined to
  be the alternative sum of the numbers of the cells of every 
  dimension, i.e., $\chi=\sum(-1)^{p}d_{p}$, where $d_{p}$ is the 
  number of the cells of $p$-dimension under a certain division 
  of the manifold. Supposing now one proceeding to take a practical 
  division of a black hole's spacetime manifold, because the observer 
  is outside the horizon, while any information inside the horizon 
  cannot escape out of the horizon, the observer can not obtain the 
  information for the division of the area inside the horizon. 
  (Meantime, we think that an observer could not enter into and out 
   again from the horizon of a black hole because of gravitation.) 
  Therefore for the observer outside the horizon, the information of 
  the division of the spacetime manifold that has sense is only of 
  that outside the horizon. That is to say 
  $d_{p}$ is the number of the cells of $p$-dimension under a certain 
  division of the manifold outside the horizon. This means that from 
  the meaning of the differential topology as in Eq. (1.3), the 
  integral area can only be taken outside the horizon.}

  According to Eq. (1.1) the differential form of degree four which is 
the Euler density reads 
$$
  \Omega=\frac{1}{32\pi^{2}}\epsilon_{abcd}\Omega^{ab}\wedge
       \Omega^{cd}~.
  \eqno{(1.4)}  $$
From Eq. (1.3) $\Omega$ is equal to the exterior derivative of a 
differential form $\Pi$ of degree three. 
To apply the Stokes theorem the integral of Eq. (1.3) can be carried 
out by the integral of $\Pi$ on the manifold's boundaries. Then for a 
four-dimensional black hole we obtain
$$
  \chi=\int_{\partial M}\Pi
      =\int_{\infty}\Pi+\int_{r_{h}}\Pi~.
  \eqno{(1.5)}  $$ 
In Eq. (1.5) the index $\infty$ means the boundary at infinity, $r_{h}$ 
means the boundary at the horizon. (For a charged or a rotating black 
hole with two horizons, here and in the following we mean the horizon 
to be the outer horizon and take this for granted.) They are 
three-dimensional hypersurfaces. 
For the topological characteristic numbers of manifolds with boundaries 
generally one should consider certain boundary corrections. The index 
theorem is called the Atiyah-Patodi-Singer index theorem generally [9]. 
For the Euler number of a black hole's spacetime manifold, 
Eguchi, Gilkey, and Hanson [9] has given a boundary modification and 
the Gauss-Bonnet formula is given by 
$$
  \chi=\frac{1}{32\pi^{2}}\int_{V}\epsilon_{abcd}\Omega^{ab}\wedge
       \Omega^{cd}-\frac{1}{32\pi^{2}}
       \int_{\partial M}\epsilon_{abcd}
       (2\theta^{ab}\wedge \Omega^{cd}-\frac{4}{3}\theta^{ab}\wedge
       \theta^{c}_{~e}\wedge\theta^{ed})~, 
  \eqno{(1.6)}  $$
where $\theta^{ab}=\omega^{ab}-(\omega_{0})^{ab}$ is the second 
fundamental form of the boundary. This formula was used for the 
calculation of the Euler numbers of four-dimensional black holes in 
Refs. [5-8]. In this paper we will use Eqs. (1.3) and (1.5) directly to 
calculate the Euler numbers for four-dimensional black holes according 
to Chern [10]. At the end of this paper, we discuss the difference and 
connection between the method used in this paper and the method used in 
Refs. [5-8] for the calculation of the Euler numbers.

  The Euler numbers for the extremal black holes need to be treated 
especially. For extremal black holes like those of the extremal  
Reissner-Nordstr\"{o}m, Kerr and Kerr-Newman black holes, or the 
extremal black holes in the superstring theories, there are two cases 
according to the studies of many authors (see, e.g., Ref. [11]) due to 
their horizons are located at the infinities or finite $r_{h}$. If their 
horizons are located at infinities, then because the spacetime metrics are 
asymptotically flat at infinities, the two terms in Eq. (1.5) are all tend 
to be zero. Therefore the Euler numbers are zero. This is in accordance 
with their topologies to be $S^{1}\times R\times S^{2}$. If their 
horizons are located at finite $r_{h}$, then one can still use Eq. (1.5) 
to calculate their Euler numbers.

  In Sec. II, we derive 
an expression of the Euler numbers for the general form of the 
metrics of four-dimensional rotating black holes. In Sec. III, 
we calculate the Euler numbers for several cases 
that includes Kerr, Kerr-Newman, and Kerr-Sen metrics. In Sec. IV, we 
discuss some of the problems and point out that the expression (2.28) 
obtained in this paper are not universal to all of the four-dimensional 
rotating black holes. In the Appendix, we derive the surface gravity 
of four-dimensional rotating black holes that is needed in the 
calculation of Sec. II.

\section{The calculation}

\indent

  The metric of a four-dimensional rotating black hole is generally 
given by 
$$
  ds^{2}=-g_{tt}(r,\theta)dt^{2}+g_{rr}(r,\theta)dr^{2}+
         g_{\theta\theta}(r,\theta)d\theta^{2}+
         g_{\phi\phi}(r,\theta)d\phi^{2}-2g_{t\phi}(r,\theta)dtd\phi~,
  \eqno{(2.1)}  $$
where in Eq. (2.1) the signature of the metric is Lorentzian. In the 
studies of the thermodynamical and topological properties of black holes 
one often needs to consider their metrics of the Euclidean signatures. 
The Euclidean metric of a rotating black hole can be obtained through 
Wick rotating both the time and angular momentum parameter of the 
Lorentzian metric, i.e., $t\rightarrow -i\tau$, $a\rightarrow ia$, 
where $\tau$ is the imaginary time [12,13]. The Euclidean form of the 
metric (2.1) can be written as   
$$
  ds^{2}=g_{\tau\tau}(r,\theta)d\tau^{2}+g_{rr}(r,\theta)dr^{2}+
         g_{\theta\theta}(r,\theta)d\theta^{2}+
         g_{\phi\phi}(r,\theta)d\phi^{2}+
           2g_{\tau\phi}(r,\theta)d\tau d\phi
  \eqno{(2.2)}  $$
generally. For the studies of the thermodynamical properties of a black 
hole, the Euclidean form metric makes a black hole be a finite 
temperature system [14,15]. The imaginary time $\tau$ is a periodic 
coordinate and all of the physical quantities like that of the Green 
functions are periodic with respect to the coordinate $\tau$. 
For the studies of the topological properties of a black hole such as 
the Euler characteristic, because it is defined for a Riemannian 
manifold which is positive definite, one also needs to consider the  
Euclidean form metric with signature $(++++)$. Thus the spacetime of 
a black hole is regarded as a total. The imaginary time $\tau$ does not 
have the time evaluation meaning any more. It is related with the 
thermodynamical properties of a black hole.

  For the convenience of the calculation we write the metric (2.2) in 
the form
$$
  ds^{2}=[a(r,\theta)d\tau+b(r,\theta)d\phi]^{2}+c^{2}(r,\theta)dr^{2}+
         f^{2}(r,\theta)d\theta^{2}+h^{2}(r,\theta)d\phi^{2}~,
  \eqno{(2.3)}  $$
where 
$$
  a=\sqrt{g_{\tau\tau}}~,   ~~~~ c=\sqrt{g_{rr}}~,
      ~~~~ f=\sqrt{g_{\theta\theta}}~,
      ~~~~ \sqrt{b^{2}+h^{2}}=g_{\phi\phi}~.
  \eqno{(2.4)}  $$
Therefore we have 
$$
  b=-\frac{g_{\tau\phi}}{a}~, ~~~~~~
  h=\sqrt{g_{\phi\phi}-\frac{g_{\tau\phi}^{2}}{g_{\tau\tau}}}~.
  \eqno{(2.5)}  $$
The local orthogonal frame is defined as 
$$ 
  ds^{2}=g_{\mu\nu}dx^{\mu}dx^{\nu}=\eta_{ab}e^{a}e^{b}~,
  \eqno{(2.6)}  $$
where $e^{a}=e^{a}_{~\mu}dx^{\mu}$. The spin connection one-form
$\omega^{a}_{~b}$ are determined uniquely by the Cantan's 
structure equations and the torsionless metric conditions
$$  
  de^{a}+\omega^{a}_{~b}\wedge e^{b}=0~, ~~~~~~       
  \omega^{a}_{~b}=-\omega^{b}_{~a}=\omega^{a}_{~b\mu}dx^{\mu}~.
  \eqno{(2.7)}  $$
The solutions of Eq. (2.7) are
$$
  \omega^{01}=\frac{1}{c}\frac{\partial a}{\partial r}d\tau-
     \frac{1}{2c}\left(\frac{\partial b}{\partial r}+
     \frac{b}{a}\frac{\partial a}{\partial r}\right)d\phi~,
  \eqno{(2.8a)}  $$
$$
  \omega^{02}=\frac{1}{f}\frac{\partial a}{\partial \theta}d\tau-
     \left(\frac{1}{2f}\frac{\partial b}{\partial \theta}+
     \frac{b}{2af}\frac{\partial a}{\partial \theta}\right)d\phi~,
  \eqno{(2.8b)}  $$
$$
  \omega^{03}=\left(\frac{1}{2h}\frac{\partial b}{\partial \theta}-
     \frac{b}{2ah}\frac{\partial a}{\partial \theta}\right)d\theta +
     \left(\frac{1}{2h}\frac{\partial b}{\partial r}-
     \frac{b}{2ah}\frac{\partial a}{\partial r}\right)dr~,
  \eqno{(2.8c)}  $$
$$
  \omega^{12}=\frac{1}{f}\frac{\partial c}{\partial \theta}dr-
     \frac{1}{c}\frac{\partial f}{\partial r}d\theta~,
  \eqno{(2.8d)}  $$
$$
  \omega^{13}=\left(\frac{a}{2ch}\frac{\partial b}{\partial r}-
     \frac{b}{2ch}\frac{\partial a}{\partial r}\right)d\tau-
     \left(\frac{1}{c}\frac{\partial h}{\partial r}+
     \frac{b}{2ch}\frac{\partial b}{\partial r}-
     \frac{b^{2}}{2ach}\frac{\partial a}{\partial r}\right)d\phi~,
  \eqno{(2.8e)}  $$
$$
  \omega^{23}=-\left(\frac{b}{2fh}\frac{\partial a}{\partial \theta}-
     \frac{a}{2fh}\frac{\partial b}{\partial \theta}\right)d\tau-
     \left(\frac{1}{f}\frac{\partial h}{\partial \theta}+
     \frac{b}{2fh}\frac{\partial b}{\partial \theta}-
     \frac{b^{2}}{2afh}\frac{\partial a}{\partial \theta}\right)d\phi~.
  \eqno{(2.8f)}  $$
The curvature two-form $\Omega^{a}_{b}$ are determined by the
following equations
$$
  \Omega^{ab}=d\omega^{ab}+\omega^{a}_{~c}\wedge\omega^{cb}~.
  \eqno{(2.9)}  $$
Then Eq. (1.4) is evaluated to be
$$
  \frac{1}{32\pi^{2}}\epsilon_{abcd}\Omega^{ab}\wedge\Omega^{cd}
  =\frac{1}{4\pi^{2}}(\Omega^{01}\wedge\Omega^{23}+
       \Omega^{20}\wedge\Omega^{13}+\Omega^{03}\wedge\Omega^{12})~.
  \eqno{(2.10)}  $$
For the metric (2.3) we obtain 
$$ 
  \Omega^{01}\wedge\Omega^{23}+
       \Omega^{20}\wedge\Omega^{13}+\Omega^{03}\wedge\Omega^{12}=
       -d(\omega^{01}\wedge\omega^{02}\wedge\omega^{03}) 
       -d(\omega^{01}\wedge\omega^{12}\wedge\omega^{13})           $$
$$
       -d(\omega^{02}\wedge\omega^{12}\wedge\omega^{23}) 
       -d(\omega^{03}\wedge\omega^{13}\wedge\omega^{23})
       +d(\omega^{01}\wedge d\omega^{23})                          $$
$$   
       +d(\omega^{13}\wedge d\omega^{02})
       +d(\omega^{03}\wedge d\omega^{12})~.
  \eqno{(2.11)}  $$
According to Eq. (1.2) we have
$$
  d\Pi=\frac{1}{4\pi^{2}}(\Omega^{01}\wedge\Omega^{23}+
       \Omega^{20}\wedge\Omega^{13}+\Omega^{03}\wedge\Omega^{12})~.
  \eqno{(2.12)}  $$
Therefore from Eqs. (2.11) and (2.12) we can read out the solution of 
$\Pi$. 
However this is not necessary for the calculation of the Euler numbers. 
The reason is that the Euler number $\chi$ is expressed by Eq. (1.5) in 
the form of the integral on the three-dimensional hypersurfaces at the 
horizon and infinity. Therefore we only need to extract the three-form of 
$d\tau\wedge d\theta\wedge d\phi$ terms from Eqs. (2.11) and (2.12).

  To calculate the Euler number for the metric (2.2) directly is rather 
complicated. However we find that it will be simplified to make a rotating 
coordinate transformation
$$
  \phi^{\prime}=\phi-\Omega_{h} \tau~,
  \eqno{(2.13)}  $$
where $\Omega_{h}$ is the angular velocity of the outer horizon and so is
a constant. Because the Euler number is a diffeomorphism invariant 
quantity it is invariant under a diffeomorphism transformation. And we 
know that an infinitesimal rotating transformation is a diffeomorphism 
transformation and a finite rotation can be obtained through many 
infinitesimal rotating transformations. Therefore the Euler number for a 
rotating black hole is invariant under the rotating (2.13). After the 
rotating (2.13) the metric (2.2) becomes
$$
  ds^{\prime~2}=G_{\tau\tau}d\tau^{2}
         +g_{rr}dr^{2}+g_{\theta\theta}d\theta^{2}
         +g_{\phi\phi}d\phi^{\prime~2}
         +2g^{\prime}_{\tau\phi} d\tau d\phi^{\prime}~,
  \eqno{(2.14)}  $$
where
$$
  G_{\tau\tau}=g_{\tau\tau}+2g_{\tau\phi}\Omega_{h}+g_{\phi\phi}
          \Omega_{h}^{2}~,~~~~~~
  g^{\prime}_{\tau\phi}=g_{\phi\phi}\Omega_{h}+g_{\tau\phi}~.
  \eqno{(2.15)}  $$
On the horizon the three-dimensional metric is diagonal because 
$$
  g^{\prime}_{\tau\phi}\big\vert_{r=r_{h}}=0~.
  \eqno{(2.16)}  $$
The reason is that for a four-dimensional rotating black hole its angular
velocity on the horizon is 
$-g_{\tau\phi}/g_{\phi\phi}\vert_{r=r_{h}}$. We also have 
$$
  G_{\tau\tau}\big\vert_{r=r_{h}}=0~.
  \eqno{(2.17)}  $$
This can be seen clearly in the Appendix.
We can see that under the rotating coordinate transformation (2.13) 
$g_{rr}$, $g_{\theta\theta}$, and $g_{\phi\phi}$ keep unchanged. $g^{rr}$ 
is not changed either. As null surfaces the horizons 
for the metrics (2.2), (2.3), and (2.14) are determined by $g^{rr}=0$.
Therefore under the coordinate transformation (2.13) the horizons of
the black holes are not changed with respect to their original locations. 
Because on the horizon $G_{\tau\tau}=0$, the horizons are also determined 
by $G_{\tau\tau}=0$. 
On the other hand the stationary limit surfaces for the metric (2.14) 
are determined by $G_{\tau\tau}=0$, it is the same equation to determine 
the horizons. Therefore for the metric (2.14) the stationary limit 
surfaces are coinciding with the horizons.

  Now for the metric (2.14) it can still be incorporated in the form
of Eqs. (2.3) to (2.5), where now
$$
  a=\sqrt{G_{\tau\tau}}~,~~~~c=\sqrt{g_{rr}}~,
     ~~~~f=\sqrt{g_{\theta\theta}}~,
     ~~~~\sqrt{b^{2}+h^{2}}=g_{\phi\phi}~.
  \eqno{(2.18)}  $$
And we have 
$$
  b=-\frac{g^{\prime}_{\tau\phi}}{a}~,~~~~~~
  h=\sqrt{g_{\phi\phi}-\frac{g_{\tau\phi}^{\prime 2}}{g_{\tau\tau}}}~.
  \eqno{(2.19)}  $$
Because on the horizon $g^{\prime}_{\tau\phi}=0$ we have 
$$
  b(r_{h},\theta)=0~,~~~~
  \frac{\partial b(r_{h},\theta)}{\partial \theta}=0~,~~~~
  h(r_{h},\theta)=\sqrt{g_{\phi\phi}(r_{h},\theta)}~.
  \eqno{(2.20)}  $$
However on the horizon $\partial b(r,\theta)/\partial r$ may 
not be zero. From Eq. (1.5) the Euler number $\chi$ is expressed by the 
integral on the three-dimensional hypersurfaces at the horizon and 
infinity. The integrands are $d\tau\wedge d\theta\wedge d\phi$ three-forms 
extracted from Eqs. (2.11) and (2.12). For 
spherically and axially symmetric black holes, the spacetime metrics 
are asymptotically flat at infinities. We can see that in Eq. (1.5) the 
integrals at infinities tend to be zero. Explicit verifications for this 
fact are omitted here. Therefore only the integrals on the horizons are 
left in Eq. (1.5).

  According to Eqs. (2.8) to (2.12) and (2.20) we can write down the 
non-zero three-form of $d\tau\wedge d\theta\wedge d\phi$ terms extracted 
from $\Pi$. They are the sum of the following four parts:  
$$
  \Pi_{1}=-\frac{1}{c}\frac{\partial f}{\partial r}
   \left[ \frac{1}{c^{2}}\frac{\partial a}{\partial r}
          \frac{\partial h}{\partial r}
         -\frac{a}{8c^{2}h}\left(\frac{\partial b}{\partial r}
          \right)^{2} \right]~, ~~~~~~
  \Pi_{2}=-\frac{1}{c}\frac{\partial f}{\partial r}
   \frac{1}{f^{2}}\frac{\partial a}{\partial \theta}
   \frac{\partial h}{\partial \theta}~,             $$
$$
  \Pi_{3}=-\frac{1}{c}\frac{\partial a}{\partial r}
   \frac{\partial}{\partial \theta}\left(\frac{1}{f}
   \frac{\partial h}{\partial \theta}\right)~, ~~~~~~
  \Pi_{4}=\frac{1}{c}\frac{\partial h}{\partial r}
   \frac{\partial}{\partial \theta}\left(\frac{1}{f}
   \frac{\partial a}{\partial \theta}\right)~.
   \eqno{(2.21)}  $$
Now the Euler number is given by the following integral:
\begin{eqnarray*}
  \chi & = & \frac{1}{4\pi^{2}}\int_{0}^{\beta}d\tau\int_{0}^{2\pi}d\phi
       \int_{0}^{\pi}d\theta (\Pi_{1}+\Pi_{2}+\Pi_{3}+\Pi_{4})
       \big\vert_{r=r_{h}}       \\
       & = & \frac{\beta}{2\pi}\int_{0}^{\pi}
       d\theta (\Pi_{1}+\Pi_{2}+\Pi_{3}+\Pi_{4})
       \big\vert_{r=r_{h}}~,    
\end{eqnarray*}
$$  \eqno{(2.22)}  $$ 
where $\beta$ is the inverse temperature. 
The Euclidean time in the integral takes one period because it is the 
periodic coordinate. According to Eqs. (A5) and (2.18) 
$a=\sqrt{G_{\tau\tau}}$ is zero on the horizon, therefore we have 
$$
  \frac{\partial a}{\partial \theta}\Bigg\vert_{r=r_{h}}=0~.
   \eqno{(2.23)}  $$
Thus in Eq. (2.22) $\Pi_{2}(r_{h},\theta)=\Pi_{4}(r_{h},\theta)=0$.
Therefore we obtain 
$$
  \chi=\frac{\beta}{2\pi}\int_{0}^{\pi}
       d\theta (\Pi_{1}+\Pi_{3})\big\vert_{r=r_{h}}~.   
   \eqno{(2.24)}  $$
According to the black hole thermodynamics, the inverse temperature 
$\beta$ is equal to $2\pi/\kappa$, where $\kappa$ is the surface 
gravity on the horizon. In the Appendix we derive a formula of Eq. (A12) 
for the surface gravity for a four-dimensional rotating black hole which 
is needed in the calculation of Eq. (2.24). In the Appendix we can also 
see that the surface gravity $\kappa$ is invariant under general 
coordinate transformations, such as the rotating of Eq. (2.13). In the 
form of Eqs. (2.3), (2.18), and Eq. (A12) we have 
$$
  \kappa=\frac{1}{c}\frac{\partial a}{\partial r} 
         \Bigg \vert_{r=r_{h}}~,~~~~~~
  \beta=\frac{2\pi c}{\partial_{r}a} \Bigg \vert_{r=r_{h}}~.
  \eqno{(2.25)}  $$
To insert Eq. (2.25) into Eq. (2.24) and to move $\beta$ inside the 
integral because the surface gravity $\kappa$ is a constant on the 
horizon as well as $\beta$ we obtain 
\begin{eqnarray*}
  \chi & = & \int_{0}^{\pi}d\theta
   \frac{c}{\partial_{r}a}\left[
  -\frac{1}{c}\frac{\partial f}{\partial r}
   \left(\frac{1}{c^{2}}\frac{\partial a}{\partial r}
   \frac{\partial h}{\partial r}
  -\frac{a}{8c^{2}h}\frac{\partial b}{\partial r}
           \frac{\partial b}{\partial r}\right)
  -\frac{1}{c}\frac{\partial a}{\partial r}
   \frac{\partial}{\partial \theta}\left(\frac{1}{f}
   \frac{\partial h}{\partial \theta}\right)\right]
   \Bigg\vert_{r=r_{h}}       \\
       & = & -\int_{0}^{\pi}d\theta\left[
   \frac{1}{c^{2}}\frac{\partial f}{\partial r}
   \left(\frac{\partial h}{\partial r}
  -\frac{a}{8c^{2}h\partial_{r}a}
     \frac{\partial b}{\partial r}
     \frac{\partial b}{\partial r}\right)
  +\frac{\partial}{\partial \theta}\left(\frac{1}{f}
   \frac{\partial h}{\partial \theta}\right)\right]
   \Bigg\vert_{r=r_{h}}~.
\end{eqnarray*}
$$  \eqno{(2.26)}  $$ 
Because $1/c^{2}=g^{rr}$, the horizons of the metrics (2.3) and
(2.14) are determined by $g^{rr}(r_{h},\theta)=0$, the first two terms 
in Eq. (2.26) vanish and we have
$$
  \chi=-\int_{0}^{\pi}d\theta
   \frac{\partial}{\partial \theta}\left(\frac{1}{f(r_{h},\theta)}
   \frac{\partial h(r_{h},\theta)}{\partial \theta}\right)~,
   \eqno{(2.27)}  $$ 
where $f=\sqrt{g_{\theta\theta}}$ and $h=\sqrt{g_{\phi\phi}}$ from 
Eq. (2.20). At last we can write Eq. (2.27) in the form 
$$
  \chi=-\int_{0}^{\pi}d\theta
      \frac{1}{2\sqrt{g_{\theta\theta}g_{\phi\phi}}}\left[
      g_{\phi\phi}^{\prime\prime}
     -\frac{1}{2g_{\phi\phi}}g_{\phi\phi}^{\prime~2}
     -\frac{1}{2g_{\theta\theta}}g_{\theta\theta}^{\prime}
      g_{\phi\phi}^{\prime}\right]
     \Bigg\vert_{r=r_{h}}~,
   \eqno{(2.28)}  $$  
where the prime is the derivative with respect to $\theta$. According 
to this formula, the Euler number for a four-dimensional spherically 
symmetric or rotating black hole is determined by $g_{\theta\theta}$ and 
$g_{\phi\phi}$ in their Euclidean metric. However whether this formula 
is universal to all of the four-dimemsional rotating black holes will be 
discussed in Sec. IV.

\section{Some examples}

\indent

  In this section we will take some examples for the above derived 
formulas (2.27) and (2.28). First we take a look at the four-dimensional 
spherically symmetric black holes. Their metrics in the Euclidean form 
are 
$$
  ds^{2}=e^{2U(r)}d\tau^{2}+e^{-2U(r)}dr^{2}+
         R^{2}(r)(d^{2}\theta+\sin^{2}\theta d\phi^{2})~.
  \eqno{(3.1)}  $$ 
We can obtain that their Euler numbers are $2$ from Eq. (2.27) directly. 
This result is also held for the four-dimensional spherically symmetric 
black holes appear in the superstring theories [16-19]. For the extremal 
four-dimensional spherically symmetric black holes such as the extremal 
Reissner-Nordstr\"{o}m black hole, if their horizons are located at 
infinities along spacelike directions like that pointed out by 
Hawking {\sl et al.} [20], their Euler numbers are zero according to 
Eq. (1.5) and their entropies are also zero. If their horizons are located 
at finite $r_{h}$ like that discussed in Ref. [11], their Euler numbers 
are still $2$ according to Eq. (2.27) and their entropies are still 
$A_{h}/4$. For the extremal black holes in the superstring 
theories they are similar as that of the extremal Reissner-Nordstr\"{o}m 
black hole. If there are Yang-Mills fields in the Einstein's gravitational 
theory, for the four-dimensional spherically symmetric black holes their 
metrics may be modified to be [21,22]
$$
  ds^{2}=e^{2U(r)}e^{-2\delta(r)}d\tau^{2}+e^{-2U(r)}dr^{2}+
         R^{2}(r)(d^{2}\theta+\sin^{2}\theta d\phi^{2})~,
  \eqno{(3.2)}  $$ 
where $e^{-2\delta(r)}$ are the corrections due to Yang-Mills fields. 
We can still obtain that their Euler numbers are $2$ according to 
Eq. (2.27).

  For the Kerr black hole its metric in the Lorentzian form is
\begin{eqnarray*}
  ds^{2}= & - & \left(1-\frac{2Mr}{r^{2}+a^{2}\cos^{2}\theta}\right)
                dt^{2}
         +\frac{r^{2}+a^{2}\cos^{2}\theta}{r^{2}+a^{2}-2Mr}dr^{2}
     +(r^{2}+a^{2}\cos^{2}\theta)d\theta^{2}         \\
          & + &  \left[(r^{2}+a^{2})\sin^{2}\theta+
      \frac{2Mra^{2}\sin^{4}\theta}{r^{2}+a^{2}\cos^{2}\theta}\right]
     d\phi^{2}-
      \frac{4Mra\sin^{2}\theta}{r^{2}+a^{2}\cos^{2}\theta}dt d\phi~,
\end{eqnarray*}
$$   \eqno{(3.3)}  $$ 
where $M$ is the total mass, $a$ is the angular momentum per unit mass. 
Through Wick rotating the time and the angular momentum parameter one 
can obtain the Euclidean form metric. In Ref. [12] it is given by 
\begin{eqnarray*}
  ds^{2}= & ~ & V\left[d\tau-2\frac{aMr\sin^{2}\theta}
              {\Delta+a^{2}\sin^{2}\theta}d\phi\right]^{2}   \\
          & + & \frac{1}{V}\left[\frac{\Delta+a^{2}\sin^{2}\theta}
              {\Delta}dr^{2}+(\Delta+a^{2}\sin^{2}\theta)d\theta^{2}
               +\Delta\sin^{2}\theta d\phi^{2}\right]~,
\end{eqnarray*}
$$  \eqno{(3.4)}  $$
where
$$
  V=1-\frac{2Mr}{r^{2}-a^{2}\cos^{2}\theta}~, ~~~~~~
  \Delta=r^{2}-2Mr-a^{2}~.
  \eqno{(3.5)}  $$
To write Eq. (3.4) in the explicit form it is
\begin{eqnarray*}
  ds^{2}= & ~ & \left(1-\frac{2Mr}{r^{2}-a^{2}\cos^{2}\theta}\right)
                d\tau^{2}
         +\frac{r^{2}-a^{2}\cos^{2}\theta}{r^{2}-a^{2}-2Mr}dr^{2}
     +(r^{2}-a^{2}\cos^{2}\theta)d\theta^{2}         \\
          & + &  \left[(r^{2}-a^{2})\sin^{2}\theta-
      \frac{2Mra^{2}\sin^{4}\theta}{r^{2}-a^{2}\cos^{2}\theta}\right]
     d\phi^{2}-
      \frac{4Mra\sin^{2}\theta}{r^{2}-a^{2}\cos^{2}\theta}d\tau d\phi~.
\end{eqnarray*}
$$  \eqno{(3.5)}  $$
To compare Eq. (3.3) with Eq. (3.6) we can see that except the change 
of the signature and $t\rightarrow\tau$, the only other change is 
$a^{2}\rightarrow -a^{2}$. All other forms of the metric coefficients 
are the same. Therefore to insert $g_{\theta\theta}$ and $g_{\phi\phi}$ 
of Eq. (3.6) into Eq. (2.28) and to note that $r_{h}^{2}-a^{2}-2Mr_{h}=0$ 
on the horizon at last we obtain
\begin{eqnarray*}
  \chi & = & 4M^{2}r_{h}^{2}\int_{0}^{\pi}
         \frac{r_{h}^{2}+3a^{2}\cos^{2}\theta}
         {(r_{h}^{2}-a^{2}\cos^{2}\theta)^{3}}\sin\theta d\theta    \\
       & = & (r_{h}^{2}-a^{2})^{2}\int_{0}^{\pi}
         \frac{r_{h}^{2}+3a^{2}\cos^{2}\theta}
         {(r_{h}^{2}-a^{2}\cos^{2}\theta)^{3}} d(-\cos\theta)~.
\end{eqnarray*} 
$$  \eqno{(3.7)}  $$
Let $\cos\theta=x$, we can obtain that the result of the integral is 
$2/(r_{h}^{2}-a^{2})^{2}$. Therefore we have
$$
  \chi_{\mbox{\small \rm Kerr}}=2~.
   \eqno{(3.8)}  $$  
This result was previously obtained by Liberati and Pollifrone [6].
Here we obtain it from a different method.

  The metric of the Kerr-Newman black hole in the Lorentzian form is 
\begin{eqnarray*}
  ds^{2}= & - &
        \left(1-\frac{2Mr-Q^{2}}{r^{2}+a^{2}\cos^{2}\theta}\right)
        dt^{2}
       +\frac{r^{2}+a^{2}\cos^{2}\theta}{r^{2}+a^{2}-2Mr+Q^{2}}dr^{2}
     +(r^{2}+a^{2}\cos^{2}\theta)d\theta^{2}         \\
          & + & \left[(r^{2}+a^{2})\sin^{2}\theta+
     \frac{(2Mr-Q^{2})a^{2}\sin^{4}\theta}
     {r^{2}+a^{2}\cos^{2}\theta}\right]d\phi^{2}-
      \frac{2(2Mr-Q^{2})a\sin^{2}\theta}{r^{2}+a^{2}\cos^{2}\theta}
      dt d\phi~,
\end{eqnarray*}
$$  \eqno{(3.9)}  $$

where $M$ is the total mass, $a$ is the angular momentum per unit
mass, and $Q$ is the electric charge. Like that of the Kerr metric, 
one can obtain its Euclidean form metric to be 
\begin{eqnarray*}
  ds^{2}= & ~ &
       \left(1-\frac{2Mr-Q^{2}}{r^{2}-a^{2}\cos^{2}\theta}\right)
        d\tau^{2}
       +\frac{r^{2}-a^{2}\cos^{2}\theta}{r^{2}-a^{2}-2Mr+Q^{2}}dr^{2}
     +(r^{2}-a^{2}\cos^{2}\theta)d\theta^{2}         \\
          & + & \left[(r^{2}-a^{2})\sin^{2}\theta-
     \frac{(2Mr-Q^{2})a^{2}\sin^{4}\theta}
     {r^{2}-a^{2}\cos^{2}\theta}\right]d\phi^{2}-
      \frac{2(2Mr-Q^{2})a\sin^{2}\theta}{r^{2}-a^{2}\cos^{2}\theta}
      d\tau d\phi
\end{eqnarray*}  
$$  \eqno{(3.10)}  $$
through Wick rotating the time and the angular momentum parameter. 
The differences between Eq. (3.9) and Eq. (3.10) are the signature, 
$t\rightarrow\tau$, and $a^{2}\rightarrow -a^{2}$. All other forms of 
the metric coefficients are the same. To insert $g_{\theta\theta}$ and 
$g_{\phi\phi}$ of Eq. (3.10) into Eq. (2.28) we obtain
$$
  \chi=(2Mr_{h}-Q^{2})^{2}\int_{0}^{\pi}
        \frac{r_{h}^{2}+3a^{2}\cos^{2}\theta}
       {(r_{h}^{2}-a^{2}\cos^{2}\theta)^{3}}\sin\theta d\theta~.
   \eqno{(3.11)}  $$ 
Eq. (3.11) is exactly Eq. (3.7) except that $2Mr_{h}$ is replaced by
$2Mr_{h}-Q^{2}$. To note that on the horizon 
$r_{h}^{2}-a^{2}-2Mr_{h}+Q^{2}=0$, therefore we have 
$$
  \chi_{\mbox{\small Kerr-Newman}}=2~.
   \eqno{(3.12)}  $$
The direct calculation for the Euler number of the Kerr-Newman black 
hole seems missing from the literatures. Here we give an exact calculation 
for this problem. For the extremal Kerr and Kerr-Newman black holes, 
there are two cases due to their horizons are located at the infinities 
or finite $r_{h}$ like that of the Reissner-Nordstr\"{o}m black hole. 
If their horizons are located at infinities, their Euler numbers are 
zero according to Eq. (1.5) for the reason that their metrics are 
asymptotically flat at infinities, and their entropies are also zero. If 
their horizons are located at finite $r_{h}$, their Euler numbers are 
still calculated as the above and therefore are $2$, their entropies are 
still $A_{h}/4$ [23].

  Next we will consider a four-dimensional rotating black hole in the 
superstring theories given by Sen [1]. 
It is a solution of the classical equations of motion of the low-energy 
effective field theory in the heterotic stirng theory. The Einstein 
canonical metric in the Lorentzian signature is given by
\begin{eqnarray*}
  ds^{2}_{E}= & - & \frac{r^{2}+a^{2}\cos^{2}\theta-2mr}
         {r^{2}+a^{2}\cos^{2}\theta+2mr\sinh^{2}(\alpha/2)}dt^{2}
         +\frac{r^{2}+a^{2}\cos^{2}\theta+2mr\sinh^{2}(\alpha/2)}
          {r^{2}+a^{2}-2mr}dr^{2}      \\
       & + & [r^{2}+a^{2}\cos^{2}\theta+2mr\sinh^{2}(\alpha/2)]
             d\theta^{2}
       -\frac{4mr a\cosh^{2}(\alpha/2)\sin^{2}\theta}
     {r^{2}+a^{2}\cos^{2}\theta+2mr\sinh^{2}(\alpha/2)}dtd\phi  \\
       & + & \{(r^{2}+a^{2})(r^{2}+a^{2}\cos^{2}\theta)
         +2mr a^{2}\sin^{2}\theta+4mr(r^{2}+a^{2})
          \sinh^{2}(\alpha/2)    \\
       & ~ & ~ +4m^{2}r^{2}\sinh^{4}(\alpha/2)\}\times
          \frac{\sin^{2}\theta}
          {r^{2}+a^{2}\cos^{2}\theta+2mr\sinh^{2}(\alpha/2)}
          d\phi^{2}~.
\end{eqnarray*}
$$  \eqno{(3.13)}  $$
It describes a four-dimensional rotating black hole with mass $M$, charge 
$Q$, angular momentum $J$, and magnetic dipole moment $\mu$ given by
$$
  M=\frac{m}{2}(1+\cosh\alpha)~, ~~~~~~
  Q=\frac{m}{\sqrt{2}}\sinh\alpha~,                $$
$$
  J=\frac{ma}{2}(1+\cosh\alpha)~, ~~~~~~
  \mu=\frac{1}{\sqrt{2}}ma\sinh\alpha~.
  \eqno{(3.14)}  $$  
The dilaton field for the solution is
$$
  \Phi=-\ln\frac{r^{2}+a^{2}\cos^{2}\theta+2mr\sinh^{2}(\alpha/2)}
       {r^{2}+a^{2}\cos^{2}\theta}~.
  \eqno{(3.15)}  $$
Its string $\sigma$-model metric is given by
$$ 
  ds^{2}=e^{-\Phi}ds^{2}_{E}~.
  \eqno{(3.16)}  $$
To transform to the Euclidean signature is the same as the above cases. 
The changes are the signature, $t\rightarrow\tau$, and  
$a^{2}\rightarrow -a^{2}$. All other forms of the metric coefficients 
are the same. The horizons for the metric (3.13) are determined by 
$g^{rr}=0$. For its Euclidean signature metric we have on the horizon 
$r_{h}^{2}-a^{2}-2mr_{h}=0$. To insert $g_{\theta\theta}$ and 
$g_{\phi\phi}$ of its Euclidean signature metric into Eq. (2.28) 
we obtain finally
$$
  \chi=4m^{2}r_{h}^{2}\cosh^{4}(\alpha/2)\int_{0}^{\pi}
       \frac{r_{h}^{2}+2mr_{h}\sinh^{2}(\alpha/2)+3a^{2}\cos^{2}\theta}
       {(r_{h}^{2}+2mr_{h}\sinh^{2}(\alpha/2)-a^{2}\cos^{2}\theta)^{3}}
       d(-\cos\theta)~.
   \eqno{(3.17)}  $$  
The result of the integral is calculated to be 
$2/4m^{2}r_{h}^{2}\cosh^{4}(\alpha/2)$. Therefore we obtain 
$$
  \chi_{\mbox{\small Kerr-Sen}}=2~.
   \eqno{(3.18)}  $$  
Therefore we give an exact calculation for the Euler number of the 
Kerr-Sen metric. It is in accordence with its topology to be 
$R^{2}\times S^{2}$.

\section{Discussion}

\indent

  In this paper we derive a formula (2.28) for the Euler numbers of 
four-dimensional rotating black holes using the Gauss-Bonnet formula
through integrating the Euler density on a black hole's spacetime 
manifold outside the horizon. From this formula we 
obtain the correct results for the Euler numbers for many cases. 
For Kerr and Kerr-Newman metrics it can be verified that their Euler 
numbers are $2$. It also stands for the Kerr-Sen metric of the 
four-dimensional rotating black hole in the heterotic string theory 
with one boost angle nonzero [1]. However this not means that the 
formula (2.28) for the Euler numbers is universal for all of the 
four-dimensional rotating black holes.

  For the other known four-dimensional rotating black holes such as the 
rotating dilaton black hole solution of Ref. [24], to use Eq. (2.28) 
we can not obtain its Euler number to be $2$. Only for its
slowly rotating case - the metric (31) of Ref. [24], we can still obtain 
$\chi=2$ from Eq. (2.28). For the four-dimensional rotating heterotic 
string black holes generated from the Kerr solution with more than one 
boost angle as those of Refs. [2-4] or the Anti-de Sitter rotating black 
hole [13,25], we can not obtain the 
correct results $\chi=2$ either from Eq. (2.28). However for all these 
four-dimensional rotating black holes, their Euler numbers are $2$ because 
their topologies are $R^{2}\times S^{2}$. This means that for these 
cases we should consider proper boundary modifications like that of 
Eq. (1.6) to obtain their Euler numbers correctly.

  At the end of this paper, it is also useful to point out the difference 
between the method used in the Introduction of this paper and the method 
used in Refs. [5-8] for the calculation of the Euler numbers. In this 
paper as we asserted in the Introduction, the integral area of the 
Euler density $\Omega$ is taken to be the area outside the horizon. 
Therefore 
the manifold of a black hole is treated as a compact manifold surrounded 
by two boundaries, one is the horizon and the other lies at the infinity. 
Therefore we do not consider again the boundary corrections of the Euler 
numbers which is the second term of Eq. (1.6). In Refs. [5-8], the 
integral areas of the Euler density $\Omega$ are also include the areas 
inside the horizons, therefore there need to consider the boundary 
corrections on the horizons, which is the second term of Eq. (1.6) that 
is obtained by Eguchi, Gilkey, and Hanson [9]. In fact it is just the 
boundary corrections on the horizons that give the correct results of the 
Euler numbers in Refs. [5-8]. Therefore we think that the method used in 
this paper is equivalent to the method used in Refs. [5-8] in fact. 
Therefore we doubt that for other four-dimensional rotating black holes 
such as those of Refs. [2-4,24,25], there may need some other boundary 
corrections other than the second term of Eq. (1.6) to obtain the correct 
results of the Euler numbers.

\vskip 1.2cm

{\large {\bf APPENDIX: SURFACE GRAVITIES FOR ROTATING 
BLACK HOLES }}
 
\vskip 0.4cm

  In this appendix we give an explanation for Eqs. (2.17) and (2.25) 
for a four-dimensional rotating black hole. The metric for a 
four-dimensional rotating black hole is given by Eq. (2.2) generally 
in the Euclidean form. For the metric (2.2) there exists the 
following Killing field
$$
  \xi^{\mu}=\frac{\partial}{\partial \tau}+
       \Omega_{h}\frac{\partial}{\partial \phi}~,
  \eqno{({\rm A}1)}$$ 
where $\Omega_{h}$ is the angular velocity of the horizon which is a 
constant. Here we mean the horizon to be the outer horizon for a rotating 
black hole. Because the horizon is defined to be a null surface and 
$\xi^{\mu}$ is normal to the horizon, we have on the horizon [26]
$$
  \xi^{\mu}\xi_{\mu}\vert_{r=r_{h}}=0~.
  \eqno{({\rm A}2)}$$ 
For the metric of Eq. (2.2) we have
$$
  \xi^{\mu}\xi_{\mu}=g_{\tau\tau}+2g_{\tau\phi}\Omega_{h}
     +g_{\phi\phi}\Omega_{h}^{2}~.
  \eqno{({\rm A}3)}$$ 
For convenience we define 
$$
    G_{\tau\tau}=g_{\tau\tau}+2g_{\tau\phi}\Omega_{h}
          +g_{\phi\phi}\Omega_{h}^{2}~.
  \eqno{({\rm A}4)}$$ 
Therefore from Eq. (A2) we have 
$$
    G_{\tau\tau}\big\vert_{r=r_{h}}=0~.
  \eqno{({\rm A}5)}$$ 
To adopt Wald's symbol of Ref. [26] we write Eq. (A3) as 
$$
  \xi^{\mu}\xi_{\mu}=-\lambda^{2}
  \eqno{({\rm A}6)}$$ 
in the spacetime of the black hole where $\lambda$ is a scalar 
function and is a constant on the horizon. According to Eq. (A3), 
$\lambda^{2}=-G_{\tau\tau}$ in fact for the metric (2.2) for a 
four-dimensional rotating black hole. Let $\nabla^{\mu}$ represent 
the covariant derivative operator, thus $\nabla^{\mu}(\xi^{\nu}\xi_{\nu})$ 
is also normal to the horizon. According to Refs. [26,27] there exists a 
function $\kappa$ such that 
$$
  \nabla^{\mu}(-\lambda^{2})=-2\kappa\xi^{\mu}~,
  \eqno{({\rm A}7)}$$ 
where on the horizon $\kappa(r_{h})$ is a constant and just the horizon's 
surface gravity.

  Samely we can set up the lower index equation 
$$
  \nabla_{\mu}(-\lambda^{2})=-2\kappa\xi_{\mu}~.
  \eqno{({\rm A}8)}$$ 
The product of Eqs. (A7) and (A8) results
$$
  \nabla^{\mu}(\lambda^{2})\nabla_{\mu}(\lambda^{2})
   =-4\kappa^{2}\lambda^{2}~.
  \eqno{({\rm A}9)}$$ 
Because $\lambda^{2}$ is a scalar function, $\kappa^{2}$ is also a 
scalar function. Therefore the surface gravity is invariant under the 
general coordinate transformations, such as the rotating coodinate 
transformation of Eq. (2.13). From Eqs. (A3), (A4), (A6), (A9), and 
the axially symmetric of the metric we obtain
$$
  4\kappa^{2}G_{\tau\tau}=g^{rr}(\partial_{r}G_{\tau\tau})^{2}
         + g^{\theta\theta}(\partial_{\theta}G_{\tau\tau})^{2}~. 
  \eqno{({\rm A}10)}$$ 
Because of Eq. (A5) we have
$$
  \lim_{r\rightarrow r_{h}}\partial_{\theta}G_{\tau\tau}=0~.
  \eqno{({\rm A}11)}$$ 
Therefore to take the limit $r\rightarrow r_{h}$ in both sides of 
Eq. (A10) results
$$
  \kappa(r_{h})=\lim_{r\rightarrow r_{h}}
     \frac{\partial_{r}\sqrt{G_{\tau\tau}}}{\sqrt{g_{rr}}}~.
  \eqno{({\rm A}12)}$$ 
In Eq. (A12) the partial derivative should be taken before the limit 
because of Eq. (A5). In the integral of Eq. (2.24) we need the 
expression of Eq. (A12) for the surface gravity of a four-dimensional 
rotating black hole.

\vskip 1cm

\noindent {\large {\bf References}}

\vskip 12pt

\noindent 

[1]A. Sen, Phys. Rev. Lett. {\bf 69}, 1006 (1992), hep-th/9204046.

[2]A. Sen, Nucl. Phys. B {\bf 440}, 421 (1995), hep-th/9411187.

[3]G. T. Horowitz and A. Sen, Phys. Rev. D {\bf 53}, 808 (1996),  
hep-th/9509108.

[4]M. Cveti\u{c} and D. Youm, Phys. Rev. D {\bf 54}, 2612 (1996), 
   hep-th/9603147.

[5]G. W. Gibbons and R. E. Kallosh, Phys. Rev. D {\bf 51}, 2839 (1995), 
   hep-th/9407118.

[6]S. Liberati and G. Pollifrone, Phys. Rev. D {\bf 56}, 6458 (1997),  
hep-th/9708014.

[7]S. Liberati, Nuo. Cim. B {\bf 112}, 405 (1997); 
   G. Pollifrone, Nucl. Phys. B (Proc. Suppl.)

   ~~ {\bf 57}, 197 (1997).

[8]K. Suzuki, Phys. Rev. D {\bf 56}, 7846 (1997), hep-th/9611095.

[9]T. Eguchi, P. B. Gilkey, and A. J. Hanson, Phys. Rep. {\bf 66}, 213 
   (1980).

[10]S. Chern, Ann. Math. {\bf 45}, 747 (1944); 
    Ann. Math. {\bf 46}, 674 (1945).

[11]B. Wang and R.-K. Su, Phys. Lett. B {\bf 432}, 69 (1998), gr-qc/9807050;  
    B. Wang, R.-K.

~~~~ Su, and P. K. N. Yu, Phys. Rev. D {\bf 58}, 124026 (1998), 
gr-qc/9801075.

[12]C. J. Hunter, Phys. Rev. D {\bf 59}, 024009 (1999), gr-qc/9807010.

[13]S. W. Hawking, C. J. Hunter, and M. M. Taylor-Robinson, 
    Phys. Rev. D {\bf 59}, 064005

    ~~~~ (1999), hep-th/9811056.

[14]G. W. Gibbons and S. W. Hawking, Phys. Rev. D {\bf 15}, 2752 (1977).

[15]G. W. Gibbons and M. J. Perry, 
    Proc. R. Soc. Lond. A {\bf 358}, 467 (1978).

[16]A. Sen, Mod. Phys. Lett. A {\bf 10}, 2081 (1995), hep-th/9504147.

[17]M. Cveti\u{c} and D. Youm, Phys. Rev. D {\bf 53}, R584 (1996), 
    hep-th/9507090.

[18]G. T. Horowitz, D. A. Lowe, and J. M. Maldacena, 
Phys. Rev. Lett. {\bf 77}, 430 (1996),

~~~~ hep-th/9603195.

[19]M. Cveti\u{c} and A. A. Tseytlin, Phys. Rev. D {\bf 53}, 5619 (1996); 
Erratum-ibid. D55

~~~~ (1997) 3907, hep-th/9512031.

[20]S. W. Hawking, G. T. Horowitz, and S. F. Ross, 
Phys. Rev. D {\bf 51}, 4302 (1995),

~~~~ gr-qc/9409013.

[21]R. Bartnik and J. McKinnon, Phys. Rev. Lett. {\bf 61}, 141 (1988).

[22]P. Bizon, Phys. Rev. Lett. {\bf 64}, 2844 (1990).

[23]J. Preskill and P. Schwarz, Mod. Phys. Lett. A {\bf 6}, 2353 (1991).

[24]J. H. Horne and G. T. Horowitz, Phys. Rev. D {\bf 46}, 1340 (1992), 
    hep-th/9203083.

[25]M. M. Caldarelli and D. Klemm, Nucl. Phys. B {\bf 545}, 434 (1999), 
hep-th/9808097.

[26]R. M. Wald, {\sl General Relativity} (University
    of Chicago Press, Chicago, 1984).

[27]J. M. Bardeen, B. Carter, and S. W. Hawking, 
    Commun. Math. Phys. {\bf 31}, 161

    ~~~ (1973).

\end{document}